\documentclass[12pt]{iopart}
\usepackage{graphicx}

\begin{document}

\title{Measurement of the electromagnetic dissociation cross section of Pb nuclei
at $\sqrt{s_{\rm NN}}$~=~2.76~TeV}

\author{}

\address{C. Oppedisano for the ALICE Collaboration}
\address{INFN Sezione di Torino, Via P. Giuria 1, 10125 Torino, ITALY}
\ead{Chiara.Oppedisano@cern.ch}
\vskip -0.3 truecm

\begin{abstract}
Electromagnetic dissociation of heavy nuclei in ultra-peripheral interactions at high
energies can be used to monitor the beam luminosity at colliders. In ALICE neutrons
emitted by the excited nuclei close to beam rapidity are detected by the Zero Degree
Calorimeters (ZDCs), providing a precise measurement of the event rate.
During the 2010 Pb run, a dedicated data taking was performed triggering on
electromagnetic processes with the ZDCs. These data, combined with the results from
a Van der Meer scan, allowed to measure the electromagnetic dissociation
cross-section of Pb nuclei at $\sqrt{s_{\rm NN}}$~=~2.76~TeV.
Experimental results on various cross-sections are presented together with
a comparison to the available predictions.
\end{abstract}
\vskip -0.3 truecm



In ultra-peripheral collisions, for which the impact parameter is larger than the sum
of the two nuclear radii, the interaction occurs via electromagnetic (EM) forces.
These interactions are usually described in the framework of the Weizs\"acker-Williams
method~\cite{WW}. 
At the LHC the cross section for EM processes exceeds the hadronic cross section by
several order of magnitude
(see~\cite{Bruce} for updated predictions).  
EM processes will therefore strongly limit the beam lifetime.
In electromagnetic dissociation (EMD) interactions the nucleus is excited by the
absorption of one or two photons (mainly in the Giant Dipole Resonance region). For
heavy nuclei single neutron (1n) emission is the main decay mechanism.
The emitted neutrons have a rapidity close to the beam one, therefore the Zero Degree
Calorimeters (ZDCs) are ideal devices for their detection. 
The ALICE experiment has two neutron ZDCs (ZNs), placed on both sides relative to the
interaction point (IP), about 114~m away from it. In addition, for the present
analysis, signals from two small EM calorimeters (ZEMs) were also used.
These detectors are placed only on one side with respect to the IP, at  about 7.5~m
from it, covering the 4.8~$<\eta<$~5.7 region (see~\cite{ALICE} for a detailed
description of the ALICE ZDCs). 
%

\begin{figure}[ht] 
\resizebox{0.48\textwidth}{!} {\includegraphics{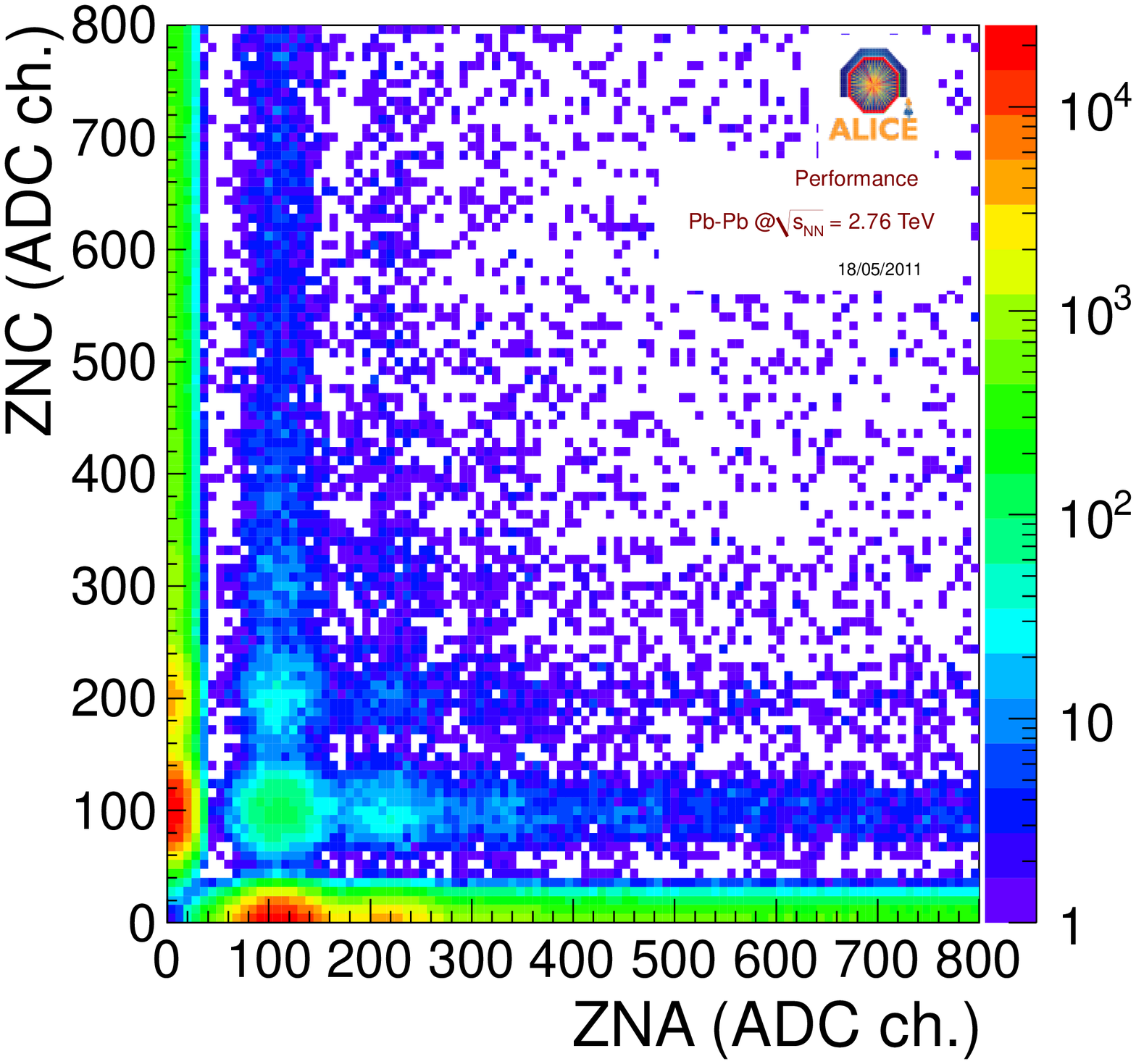}}
\resizebox{0.48\textwidth}{!} {\includegraphics{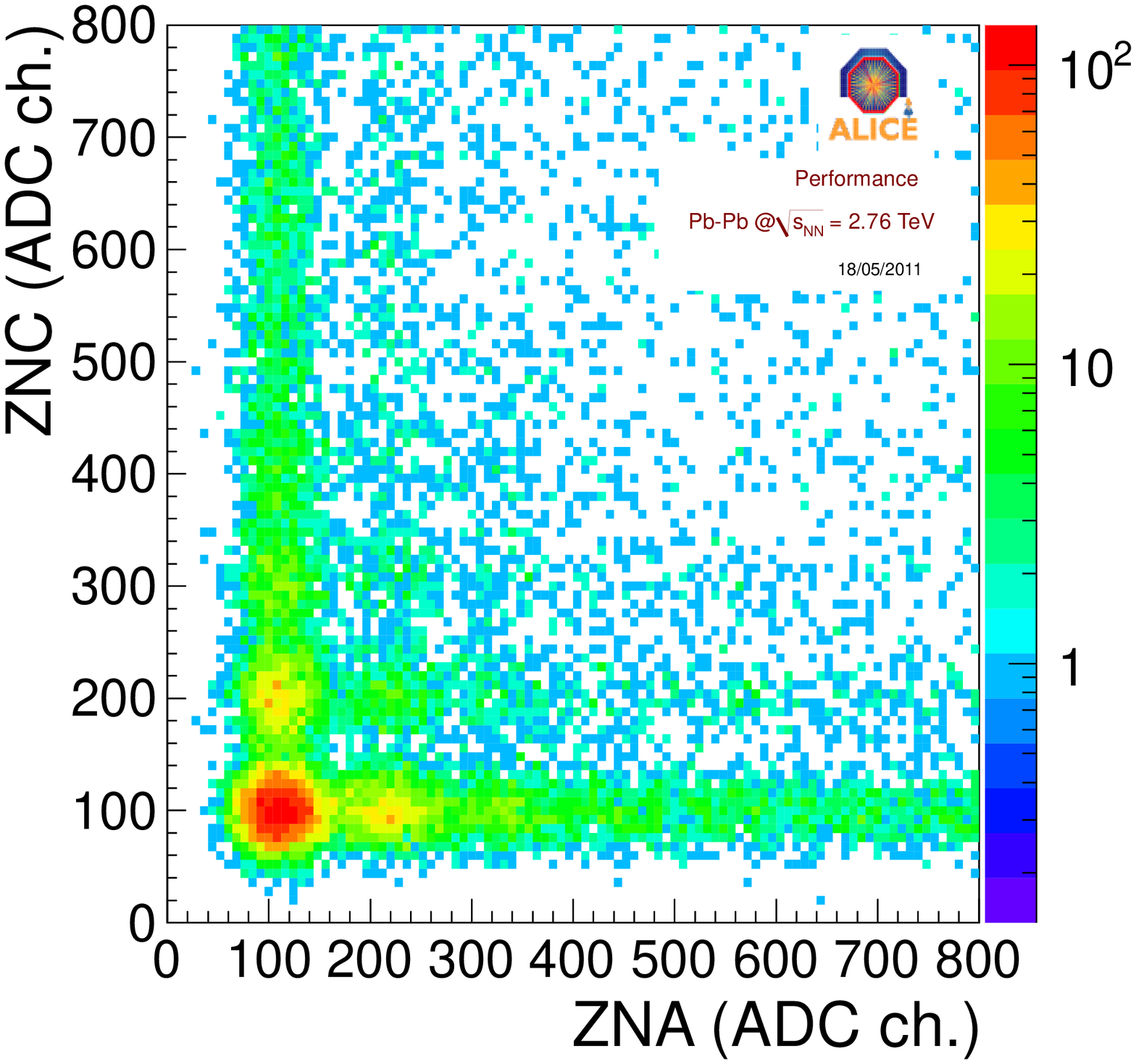}}
\vskip -0.5 truecm
\caption{\label{fig1} Left: ZNC vs. ZNA signal for single EMD events. Right: 
ZNC signal vs. ZNA signal for mutual EMD events (single neutron signal is at ADC 
channel 100).} 
\end{figure}

The analyzed data sample ($\sim$3$\times 10^{6}$ events) was collected in a
dedicated run during 2010 Pb data taking. The trigger condition required a signal
either in A side ZN (ZNA) or in C side ZN (ZNC), selecting thus single
EMD\footnote{Single EMD indicates interactions in which at least 1n is emitted by
one of the two nuclei. In mutual EMD processes both nuclei emits at least 1n (single
EMD includes therefore mutual EMD).} as well as hadronic interactions. Requiring
the coincidence of the signal in the two ZNs, only mutual EMD processes and hadronic
interactions are selected (fig.~\ref{fig1}).

Data from a van~der~Meer (vdM) scan~\cite{vdM} have also been used. During the scan 
different ZDC trigger inputs have been used to select various processes (as reported in
table~\ref{tab1}). The vdM scan results are still preliminary, in particular the
systematic error estimate. 
The main contribution to the systematic errors are: $\pm$4\% from systematic
uncertainties associated with the ALICE measurement 
(see~\cite{KO} for a complete description of the sources), 
$\pm$3\% from beam current measurement and -0\%/+11\% from the ghost
charge\footnote{The amount of charge outside the nominal buckets.} fraction evaluation.
The main contribution to the systematic error on cross section calculation is therefore
presently due to the error on absolute cross section measurements from the vdM  scan.
\begin{table}
\caption{\label{tab1}Preliminary results from the van~der~Meer scan for three ZDC
trigger inputs.}
\footnotesize\rm
\begin{tabular*}{\textwidth}{@{}l*{15}{@{\extracolsep{0pt plus12pt}}l}}
\br
ZDC trigger&Process&$\sigma_{vdM} (b)$\\
\mr
ZNC OR ZNA&2$\times$single EMD - mutual EMD& 362.61 $\pm$ 0.02~stat. $^{+39.89}_{-10.88}$ syst.\\
 & + hadronic&\\
(ZNC AND ZNA) AND& &\\
NOT (ZEM1 OR ZEM2)&mutual EMD& 5.91 $\pm$ 0.18~stat. $^{+0.65}_{-0.18}$ syst.\\
(ZNC AND ZNA) AND& &\\
(ZEM1 OR ZEM2) &hadronic& 7.08 $\pm$ 0.15~stat. $^{+0.78}_{-0.21}$ syst.\\
\br
\end{tabular*} 
\vskip -0.5 truecm
\end{table}

The model used for comparison is the Relativistic ELectromagnetic DISsociation
(RELDIS)~\cite{RELDIS0, RELDIS}, in which the EM dissociation of
ultra-relativistic heavy ions is treated including both single and double virtual
photon absorption. The EMD cross section predictions for Pb-Pb interactions at
$\sqrt{s_{\rm NN}}$~=~2.76~TeV are:  $\sigma^{sEMD}$~=~(185.2~$\pm$~9.2)~b for single
EMD,  $\sigma^{mEMD}$~=~(5.5~$\pm$~0.6)~b for mutual EMD processes.  
The error on the calculated single EMD cross section also includes the systematic error
estimated as the difference between the RELDIS result and the calculations by other
authors~\cite{Baltz}.
%

\begin{figure}[ht] 
\resizebox{0.48\textwidth}{!} {\includegraphics{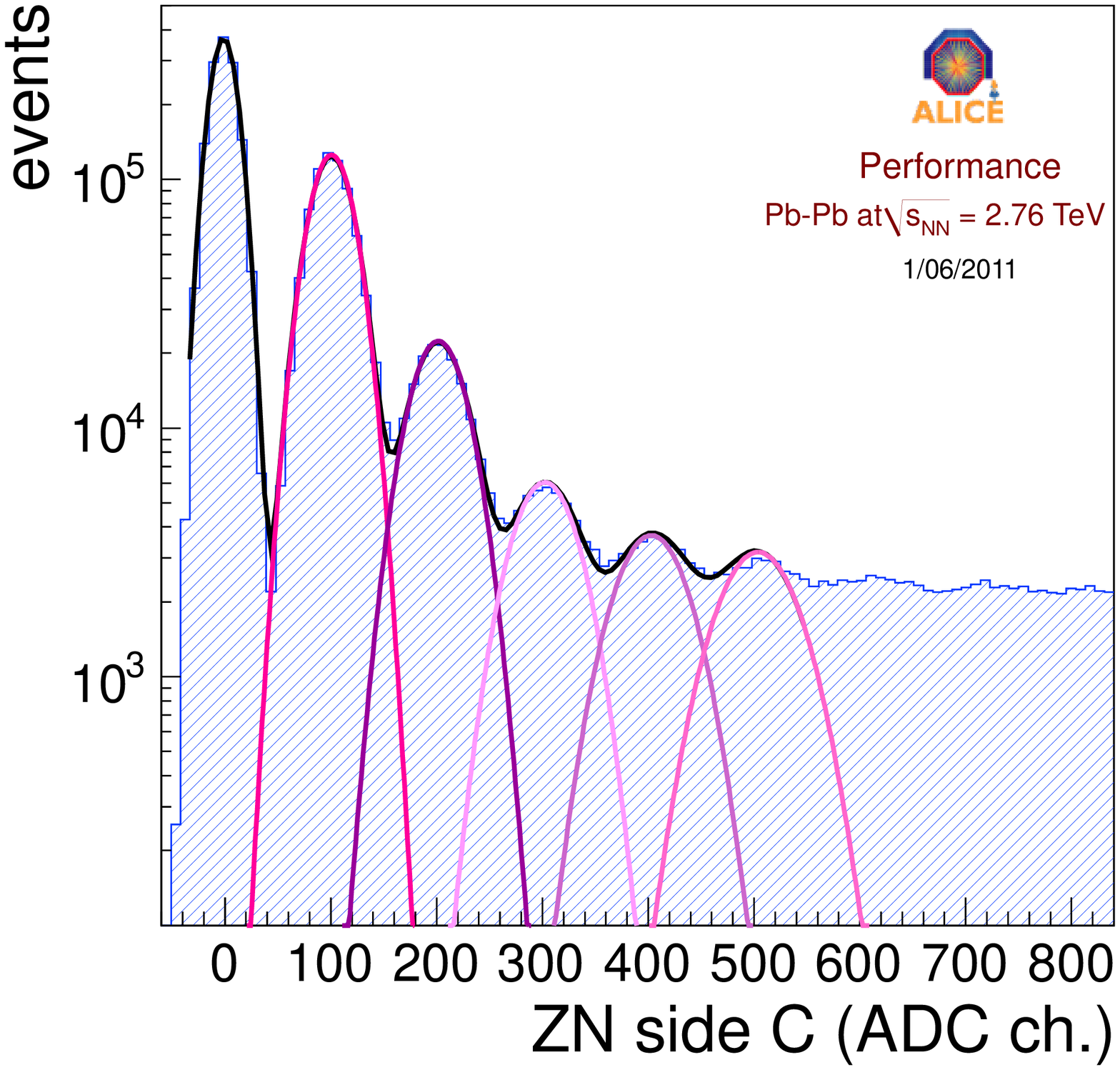}}
\resizebox{0.48\textwidth}{!} {\includegraphics{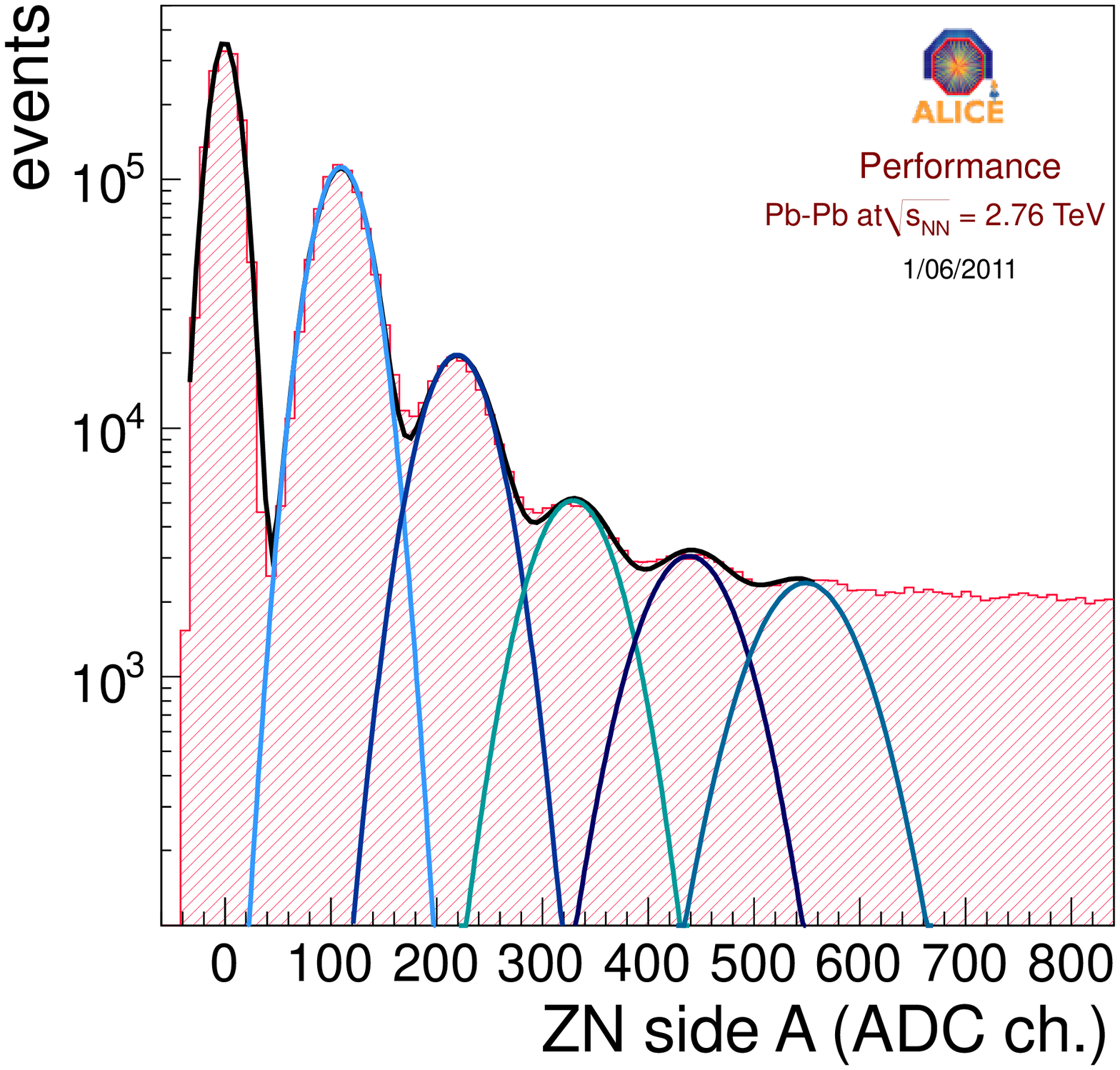}}
\vskip -0.5 truecm
\caption{\label{fig2.0} ZNC (left) and ZNA (right) spectra (event selection: ZNC or ZNA).} 
\end{figure}

Fig.~\ref{fig2.0} shows the two ZN signal amplitude spectra. The superimposed fit is
given by the sum of 6 Gaussians. The curves for the pedestal  (corresponding to neutron
emission on the other side) and for the 1n peak have three free parameters, while the
following Gaussians have a constraint both on the mean value (fixed to be $i$ times the
1n mean value for i$^{th}$ peak) and on the width (to take into account that the
pedestal width affects the width of the Gaussian signals).

\begin{figure}[ht] 
\resizebox{0.48\textwidth}{!} {\includegraphics{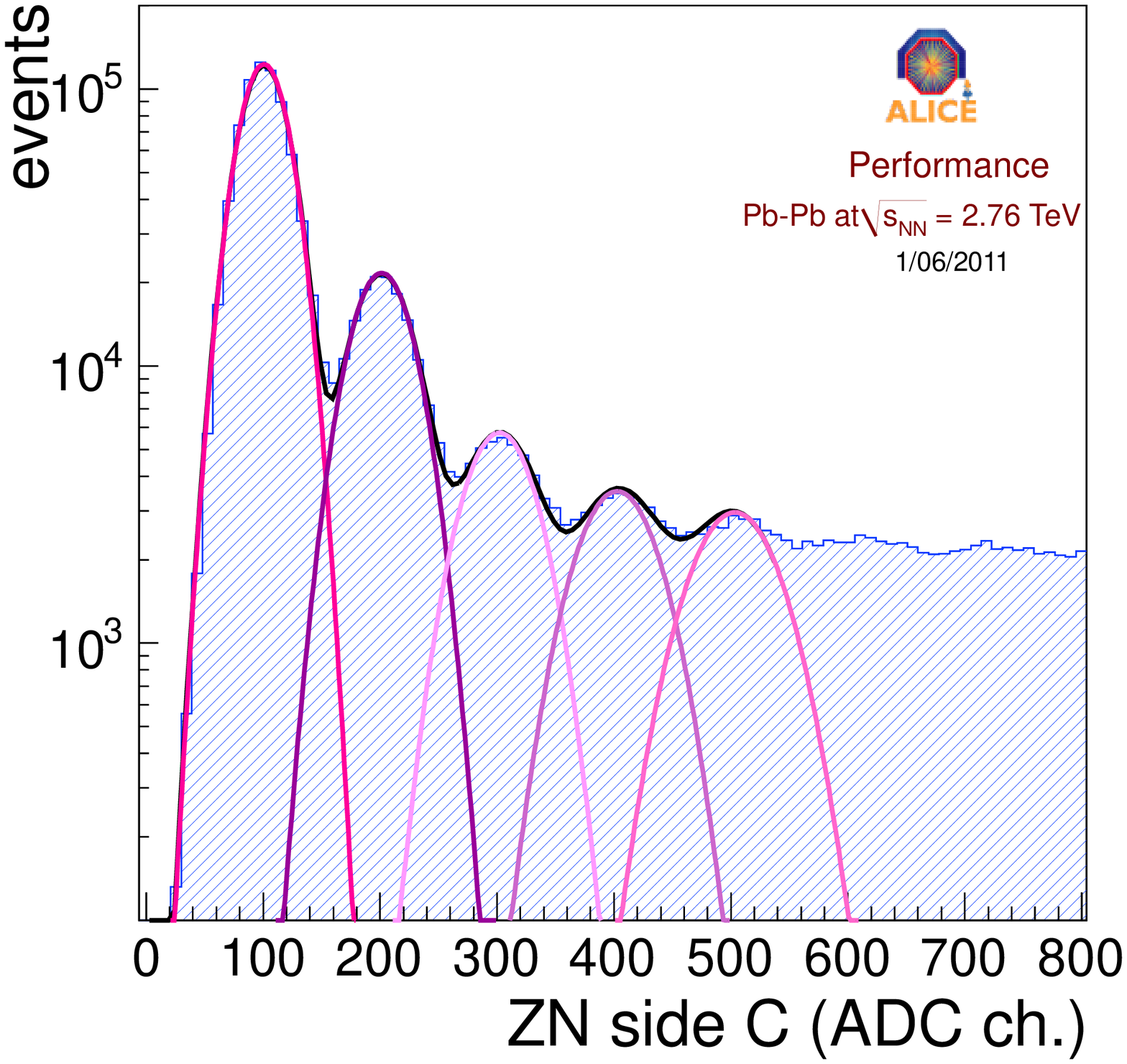}}
\resizebox{0.48\textwidth}{!} {\includegraphics{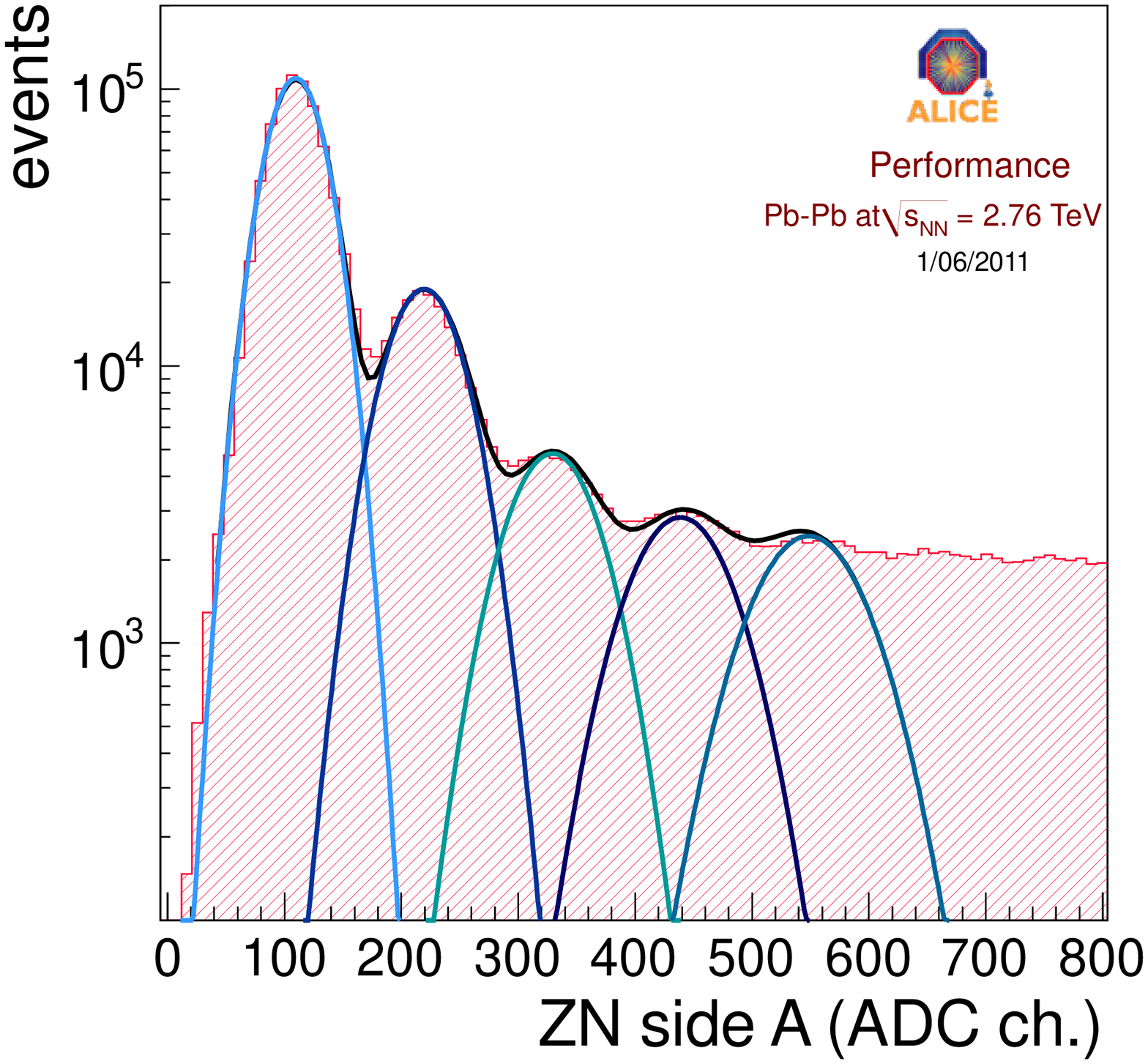}}
\vskip -0.5 truecm
\caption{\label{fig2} ZNC (left) and ZNA (right) spectra (event selection: signal
in one ZN but not in the other one).} 

\end{figure}

For hadronic interactions, both ZNs have a signal above threshold\footnote{Threshold
values are estimated to be $\sim$500~GeV for ZNC and  $\sim$450~GeV for ZNA.}.
Therefore,  requiring to have a signal above threshold in one of the two ZN and no
signal in the other one, hadronic events are rejected. The spectra with such an event
selection are shown in fig.~\ref{fig2} together with the fit obtained by summing five
Gaussians, similarly to what described above.
For both selections the cross section for the selected process can be estimated using 
the absolute values measured in the vdM scan: 
$\sigma^{proc} = \sigma_{vdM} \times N^{proc}/N_{tot}$, where $N^{proc}$ is
the number of events in our sample for the selected process and $N_{tot}$ is the
total number of events collected with the same trigger used to determine 
$\sigma_{vdM}$. 
Finally the calculated values are corrected for ZN acceptance\footnote{The acceptance
for neutrons emitted in EMD processes at $\sqrt{s_{\rm NN}}$~=~2.76~TeV, evaluated
through a Monte Carlo using RELDIS as input, is $\sim$99\%.}.

The mutual EMD cross section calculation is much more delicate since its value is
comparable to the hadronic cross section value and moreover the event selection
(reported in table~\ref{tab1}) depends on the ZEM acceptance.  Through Monte Carlo
simulation, using RELDIS as input, we estimated that 97\% of the events fulfill the
chosen event selection for mutual EMD events (i.e., give no signal in ZEM). Using
HIJING~\cite{HIJING} as event generator, in 92\% of the cases a minimum bias hadronic
interaction gives a signal over threshold in one of the two ZEM\footnote{The energy
threshold is fixed to the experimental value: $\sim$15~GeV for each ZEM.}.
Applying these correction factors, we estimated the contamination of hadronic
events in the mutual EMD sample and then we calculated the related cross section 
value.
The results are summarized and compared to the predictions in table~\ref{tab3}.
\begin{table}
\caption{\label{tab3}Cross section measurements and predictions from
RELDIS~\cite{RELDIS}.}
\footnotesize\rm
\begin{tabular*}{\textwidth}{@{}l*{15}{@{\extracolsep{0pt plus12pt}}l}}
\br
Process&Data $\sigma$ (b)&RELDIS $\sigma$ (b)\\
\mr
$\sigma^{sEMD}+\sigma^{had}$ & 195.6 $\pm$ 0.1 stat. $^{+24.2}_{-11.7}$ syst.& 192.9 $\pm$ 9.2\\
$\sigma^{sEMD}-\sigma^{mEMD}$ & 176.9 $\pm$ 0.1 stat. $^{+21.6}_{-10.6}$ syst.& 179.7 $\pm$ 9.2\\
$\sigma^{mEMD}$ & 5.7 $\pm$ 0.2 stat. $^{+0.7}_{-0.3}$ syst.& 5.5 $\pm$ 0.6\\
$\sigma^{sEMD}$ & 185.7 $\pm$ 0.2 stat. $^{+22.6}_{-11.1}$ syst.& 185.2 $\pm$ 9.2\\
\br
\end{tabular*}

\vskip -0.3 truecm

\end{table}


Cross section values for EMD processes have been measured in Pb-Pb collisions at
$\sqrt{s_{\rm NN}}$~=~2.76~TeV detecting the emitted neutrons and using the absolute
cross section values measured in a van der Meer scan. An excellent agreement with the
predictions from the RELDIS model is found.  We conclude that the ALICE ZDCs can
provide a robust and independent way to monitor the LHC luminosity in heavy ion
collisions by measuring the rate of emitted neutrons.

\end{document}